\documentclass{aa}
\usepackage{graphicx}
\usepackage{natbib, amssymb, amsmath}
\bibpunct{(}{)}{;}{a}{}{,}
\begin{document}
\def\soo{SO$_2$}
\def\rcm{cm$^{-1}$}
\def\ccms{cm$^{-3}$~s$^{-1}$}
\def\gtsim{{_>\atop{^\sim}}}
\def\ltsim{{_<\atop{^\sim}}}
\title{An atomic and molecular database for analysis of submillimetre line
observations}

\author{F. L. Sch{\"o}ier\inst{1,3}  \and 
        F. F. S. van der Tak\inst{2} \and
        E. F. van Dishoeck\inst{3} \and
        J. H. Black\inst{4}}

\offprints{F. L. Sch\"oier \\ \email{fredrik@astro.su.se}}

\institute{ Stockholm Observatory, AlbaNova, SE-106 91 Stockholm, Sweden 
\and Max-Planck-Institut f{\"u}r Radioastronomie, Auf dem H{\"u}gel 69, 53121 Bonn, Germany
\and Leiden Observatory, P.O. Box 9513, NL-2300 RA Leiden, The Netherlands 
\and Onsala Space Observatory, SE-439 92 Onsala, Sweden}

   \date{Received; accepted}

   \abstract{Atomic and molecular data for the transitions of a
     number of astrophysically interesting species are summarized,
     including energy levels, statistical weights, Einstein
     $A$-coefficients and collisional rate coefficients.  Available
     collisional data from quantum chemical calculations and
     experiments are extrapolated to higher energies (up to $E/k \sim 1000$ K).
     These data, which are made publically available through the WWW at
     {\tt http://www.strw.leidenuniv.nl/$\sim$moldata},
     are essential input for non-LTE line radiative transfer
     programs.   An online version of a computer program for performing statistical
     equilibrium calculations is also made available as part of the database.
     Comparisons of calculated emission lines using different sets of
     collisional rate coefficients are presented.
     This database should form an important tool in analyzing
     observations from current and future (sub)millimetre and infrared 
     telescopes.

   }

   \maketitle

\section{Introduction}
A wide variety of molecules has been detected in space to date
ranging from simple molecules like CO to more complex organic
molecules like ethers and alcohols.  Observations of molecular lines
at millimetre and infrared wavelengths, supplemented by careful and
detailed modelling, are a powerful tool to investigate the physical
and chemical conditions of astrophysical objects
\citep[e.g.,][]{Genzel91, Black00}. 
To constrain these conditions,
lines with a large range of critical densities and excitation
temperatures are needed, since densities typically range from
$\sim10^2-10^9$\,cm$^{-3}$ and temperatures from $\sim 10-1000$\,K in
the interstellar and circumstellar environments probed by current and
future instrumentation.

In recent years, different molecules have been developed as tracers
for different  physical and
chemical conditions (see \nocite{Dishoeck99}van Dishoeck \& Hogerheijde 1999 for a review). 
For example, CO is used as a tracer of the
total gas mass whereas readily observed molecules with large dipole
moments, such as CS, HCO$^{+}$ and HCN constrain the density
structure.
The wide variety of H$_2$CO and CH$_3$OH lines accessible at
millimetre and submillimetre wavelengths trace both the temperature
and density structure \citep[e.g.,][]{Mangum93}. Organic
molecules like CH$_3$OCH$_3$ and CH$_3$CN probe the chemical
complexity.  Deuterated molecules contain a record of
the conditions and duration of the cold pre-stellar phase.  Si- and
S-bearing molecules, in particular SiO and SO$_2$, probe shocks. Lines
of the main species as well as the (generally) optically thin isotopes
are needed to determine accurate abundances and line profiles.  High
frequency lines and vibrationally excited lines are particularly
valuable for probing the warm and dense inner parts of the
circumstellar envelopes \citep[e.g.,][]{Ziurys86, Boonman01}.  

To extract astrophysical parameters, the excitation and radiative
transfer of the lines need to be calculated.  Indeed, it is becoming
increasingly clear that more information  --- including chemical gradients
throughout the source --- can be inferred from the
data if a good molecular excitation model is available
\citep[e.g.,][]{Schoeier02,Maret04}. The simplest models adopt the
`local' approximation, for example in the widely used large velocity
gradient (LVG) method.  A number of more sophisticated, non-local radiative
transfer codes have been developed for the interpretation of molecular
line emission \citep[e.g.,][see \nocite{Zadelhoff02}van Zadelhoff et al.\ 2002 for a review]{Bernes79, Juvela97, Hogerheijde00,
Ossenkopf01, Schoeier01}.  The
application of these codes ranges from protostellar environments to
the circumstellar envelopes of late-type stars.  The radiative
transfer analysis requires accurate molecular data in the form of energy
levels, statistical weights and transition frequencies as well as the
spontaneous emission probabilities and collisional rate coefficients.
The JPL\footnote{\tt{http://spec.jpl.nasa.gov}} catalog
\citep{Pickett98},
HITRAN\footnote{\tt{http://cfa-www.harvard.edu/HITRAN/}} database
\citep{Rothman03}, and the CDMS\footnote{\tt{http://www.cdms.de}}
catalogue \citep{Muller01} contain energy levels and transition
strengths for a large number of molecular species. Detailed summaries
of the theoretical methods and the uncertainties involved in
determining collisional rate coefficients are given by
\citet{Green75}, \citet{Roueff90} and \citet{Flower90}.  In this
paper, these and other literature data on the rotational transitions
of 23 different molecules are summarized and extrapolations of
collisional rate coefficients to higher energy levels and temperatures
are made. The molecular data files can be found at the webpage
{\tt http://www.strw.leidenuniv.nl/$\sim$moldata} and is the first effort
to systematically collect and present the data in a form easily used
in radiative transfer modelling of interstellar regions.  The focus is
on rotational transitions within the ground vibrational state, but the
lowest vibrational levels are included for a few common species where such
data are available. Many of the data files presented here were adopted
by \citet{Schoeier02} to model the circumstellar environment of the
protostar \object{IRAS 16293--2422}. In addition, data files for three atomic species
are presented. 
The excitation of atomic fine structure levels plays an important role in cooling of a wide variety of astrophysical objetcs.

 An online version of RADEX\footnote{\tt{http://www.strw.leidenuniv.nl/$\sim$moldata/radex.html}}, a statistical equilibrium radiative transfer code using an escape probability formalism,
is made available for public use as part of the database. RADEX is
comparable to the LVG method and provides a useful tool for rapidly
analysing a large set of observational data providing constraints on
physical conditions, such as density and kinetic temperature
\citep{Jansen94,Jansen95}.  RADEX provides an alternative to the widely used
rotation temperature diagram method \citep[e.g.,][]{Blake87} which
relies upon the availability of many optically thin emission lines and
is useful only in roughly constraining the excitation temperature in
addition to the column density.
  A guide for using the code in practice is provided
at the RADEX homepage.
RADEX will be presented in more detail in a forthcoming paper (van der Tak et al., in prep.) at which point the source code will be made publically available.

\section{Energy levels}
In this section the molecular structure is briefly reviewed.
This serves merely to provide 
some basic information needed to properly use the data files. 
Detailed
discussions on molecular (and atomic) structure can be found in, e.g.,
\citet{Townes75}.

%
%

%

\subsection{General considerations}
The energy levels are obtained from the JPL, HITRAN,
and CDMS catalogues.  The energy levels and the corresponding line
frequencies are thus of spectroscopic quality and may be used for the
purpose of line identification,  unless stated otherwise.


Generally, we retain only the ground vibrational state and include
energy levels up to $E/k \sim 1000$ K. Vibrationally excited levels are
usually not well populated in the regions probed by current
(sub)millimetre telescopes.  Moreover, little is known about
collisional rate coefficients for vibrational transitions \citep[e.g.,][]{Chandra01}. 
However, for some specific molecules, e.g.\ CO and CS, vibrationally excited
levels are also included. In the prolongation of this project, data
files including vibrational levels will be added for more molecular
species.

Molecules with ortho and para versions (or $A$- and $E$-type as in the
case of e.g.\ CH$_3$OH) are treated as separate species.

\subsection{Linear molecules}
The energy levels for diatomic and linear polyatomic molecules in the
$^1\Sigma$ electronic state are quantified, to first order, according
to
\begin{equation}
\label{linear}
E= BJ(J+1),
\end{equation}
where $B$ is the rotational constant and related to the moment of
inertia $I$, around axes perpendicular to the internuclear axis,
through $B=(2I)^{-1}$.  Heavy linear molecules, like HC$_3$N, 
have more densely spaced energy levels than diatomic molecules like,
e.g., CO.  These pure rotational energy levels are classified
according to the rotational quantum number $J$ and their statistical
weights are
\begin{equation}
\label{linear_weight}
g=(2J+1).
\end{equation}
Note that to obtain state energies of
spectroscopic accuracy, Eq.~(\ref{linear}) must be augmented with centrifugal
distortion ($\propto J^2(J+1)^2$) and higher-order terms.
The majority of molecular species presented here have a $^1\Sigma$
electronic ground state, i.e., the sum of the orbital angular momenta
of their electrons and the sum of the electron spins are 
both zero. However, there are some exceptions where either can be
non-zero.

For a molecule in a $^2\Sigma$ electronic ground state, e.g., SO and
CN, the sum of the electron spins is 1/2. The non-zero spin
creates a splitting of the levels due to coupling between the electron
spin and the total angular momentum of the molecule. The total angular
momentum is quantified according to $N$ and includes the rotation of
the molecule. Molecules like, e.g., O$_2$ have a total electron
spin of 1 in a $^3\Sigma$ electronic ground state.
Some important molecules such as NO, NS, and OH have a $^2\Pi$ ground state
with a total electronic orbital momentum of 1 and total spin of 1/2.
Spectroscopically, such molecules show `$\Lambda-$doubling', with
$^2\Pi_{1/2}$ and $^2\Pi_{3/2}$ ladders.

The various molecular angular momenta may couple together in many
different ways, such as spin-orbit and spin-spin coupling. Ideally,
these fall in one of five different classes, known as Hund's coupling
cases. In practice, intermediate cases often occur; see
\citet{Townes75} for details.

\subsection{Non-linear molecules}
The structure of non-linear molecules, such as e.g.\ H$_2$CO
and CH$_3$OH, is more complex. Rotation can take place around
different axes of inertia, characterized by the rotational constants
$A$, $B$ and $C$ which, in absence of any symmetry, involve different
amounts of energy, $A$$>$$B$$>$$C$.  The degree of asymmetry is
measured by Ray's parameter
\begin{equation}
\label{ray}
\kappa = \frac{2B-A-C}{A-C},
\end{equation}
and is $-$1 for a prolate symmetric top ($B$$=$$C$, e.g. CH$_3$CN) and
$+$1 for an oblate symmetric top ($B$$=$$A$, e.g. NH$_3$). 
Asymmetric rotors such as H$_2$O have $|\kappa|\ll1$.

The energy levels of symmetric top molecules, such as NH$_3$ and
CH$_3$CN, are described by the quantum numbers $J$ and $K$, where $K$
is the projection of the total angular momentum $J$ on the symmetry
axis. For a prolate symmetric top molecule, the energy
of a rotational level is given (to first order) by
\begin{equation}
\label{symrotor}
E = BJ(J+1) + (A-B)K^2.
\end{equation}

The energy levels for a slightly asymmetric prolate top such as H$_2$CO
can be calculated from
\begin{equation}
\label{rotor}
E = \frac{B+C}{2} J(J+1) + \left( A - \frac{B+C}{2} \right) w_{p},
\end{equation}
where $w$$\approx$$K^2$ with corrections due to the slight asymmetry
[\citet{Townes75}, Appendix III].

\subsection{Hyperfine splitting}
A further complication arises when the nuclear spin couples to the
rotation producing what is known as hyperfine splitting. The
astrophysically most relevant cases are when the molecule contains a
$^{14}$N or D nucleus. When the lines are spectroscopically resolved, hyperfine
structure provides information on the optical depths, which
is otherwise hard to obtain (e.g., \citealt{schmid04}).

 Hyperfine splitting can be important in line transfer and
introduce non-local effects for lines overlapping in frequency, e.g., the $J=1\rightarrow 0$ lines of
N$_2$H$^+$ and HCN. The common assumption is that the hyperfine components each have the
same excitation temperature.  However, exceptions to this rule
(`hyperfine anomalies') have been observed and more detailed
treatments developed \citep{stutzki85,Truong89,Lindqvist00}.
 Often the splitting
between individual hyperfine components is small, producing lines
which are separated in frequency by a small amount compared to the
line-broadening, so that this splitting can be safely neglected and
treated as a single level for the purpose of excitation analysis. 

The first release of the database includes hyperfine splitting for some of the most 
relevant  molecules, such as HCN and OH.  Future releases
will present data files with hyperfine splitting included for additional species.

\section{Radiative rates}
\subsection{General formulae}
The radiative rates for dipole transitions from an upper state $u$ to
a lower state $l$ can be calculated from
\begin{equation}
A_{ul} = \frac{64 \pi^4 \nu^3 \mu^2}{3c^3 h}\frac{\mathcal{S}_{ul}}{g_u},
\end{equation}
where $\mu$ is the electric dipole moment and $\mathcal{S}$ is the
transition strength. The transition strength depends on the complexity
of the molecule and is explained below in some detail.  Strictly
speaking, the dipole moment should be averaged over the vibrational
wavefunction(s) of the transition involved ($\mu_v$), but in practice
often the dipole moment appropriate for the equilibrium geometry is
taken ($\mu_e$). The electric dipole moments are assumed to be the
same for all isotopes of a particular molecule, even though small
differences exist for $\mu_v$. Because of the $\nu^3$ factor, the
resulting Einstein $A$-values can still differ considerably for
isotopes, especially for deuterated species.

%
For transitions with $\Delta J$$\pm$1 in linear molecules, the transition strength is
\begin{equation}
\mathcal{S}_{ul} = g_l
\end{equation}
whereas for symmetric top molecules, the transition strength from level $J,K$ to $J-1,K$ is given by
\begin{equation}
\mathcal{S}_{ul} = \frac{J^2-K^2}{J}.
\end{equation}
In the general case of asymmetric tops, simple expressions for
$\mathcal{S}$ do not exist.  

\subsection{Dipole moments}
\label{dipole}
Transition strengths are available from the spectroscopic databases
mentioned above (JPL, CDMS, HITRAN).  There is some inconsistency in
the astrophysical literature regarding the choice of values of
electric dipole moments, however. This often manifests itself as an
apparent bias against results of {\it ab initio} theoretical
calculations, even when experimental results for transient species are
merely estimated or wholly absent.  A case in point concerns the pair
of ions HCO$^+$ and HOC$^+$: the widely cited JPL catalogue 
offers $\mu=3.30$ D
for HCO$^+$ and $\mu=4.0$ D for HOC$^+$ based on low-level theoretical
estimates of \citet{Woods75}  and \citet{Gudeman82},
respectively, whereas accurate {\it ab initio} values from
\citet{Botschwina93} give $\mu_0$(HCO$^+$) = $3.93\pm .01$ D and
$\mu_e$(HOC$^+$) = 2.74 D. 
\citet{Ziurys95} adopted a similar value,
$\mu_0$(HOC$^+$) = 2.8 D, from an {\it ab initio} computation of
\citet{DeFrees82}. Because the inferred column densities scale as
$\propto \mu^{-2}$, these discrepancies in dipole moments can result
in errors of factors of two in derived abundances.  

Table~\ref{tab_dipole} collects values of dipole moments for a
(non-exhaustive) sample of molecules of astrophysical interest.  Users
are encouraged to remain aware of the original literature.
Unless otherwise indicated, all entries refer to the electronic and 
vibrational ground states. 

For small dipoles, centrifugal corrections to the dipole moment are
appreciable. In the case of CO, rotational effects reduce the
$A$--value by 1\% for $J$=7 and by 10\% for $J$=22. The JPL and CDMS
catalogues consider this effect and so do our datafiles.

  \begin{table}
      \caption[]{Summary of adopted dipole moments$^{a,b}$
      }
 \label{tab_dipole} 
 $$
         \begin{array}{p{0.12\linewidth}cccc}    
    \hline
    \noalign{\smallskip}
    \multicolumn{1}{c}{\mathrm{Molecule ^{\mathrm a}}} & 
            \multicolumn{1}{c}{\mu_0} &
    \multicolumn{1}{c}{\mu_e} &
    \multicolumn{1}{c}{\mathrm{Method}} &
    \multicolumn{1}{c}{\mathrm{Reference}} 
     \\
             &
             \multicolumn{1}{c}{\mathrm{[D]}} & 
    \multicolumn{1}{c}{\mathrm{[D]}} &
            & 
            
             \\

    \noalign{\smallskip}
            \hline
 CO & 0.110 & & {\mathrm{expt.}} & 1 \\
 SO & 1.52\pm .02 & & {\mathrm{expt.}} & 2  \\
SO$_2$ & 1.633 & & {\mathrm{expt.}} &  3 \\
CS &  1.958\pm .005 & & {\mathrm{expt.}} &  4\\
SiO &  3.098 & &{\mathrm{expt.}} &  5 \\
SiS & 1.73\pm .06 & & {\mathrm{expt.}} & 6 \\
 HCO$^+$ & -3.93\pm .01 & -3.90\pm .01 & {\it ab\ initio} & 7  \cr
 HOC$^+$ & 2.74 &  2.8 & {\it ab\ initio} & 8, 9  \cr
 OCS &  0.7152 & & {\mathrm{expt.}} & 10  \\
HC$_3$N & 3.732 && {\mathrm{expt.}} & 11 \cr
 HCN & 2.985 && {\mathrm{expt.}} & 12 \cr
 HNC & 3.05\pm .1 && {\mathrm{expt.}} & 13\cr
 c-C$_3$H$_2$ & 3.27\pm .01 & & {\mathrm{expt.}} & 1 \cr
CH$_3$CN & 3.922 & & {\mathrm{expt.}}  &  14 \\
H$_2$CO &  2.332 & &{\mathrm{expt.}} & 15 \\
 N$_2$H$^+$ & &3.4 &{\it ab\ initio} &  16\cr
HCS$^+$ & 1.958  & & {\it ab\ initio} &  17\\
CH$_3$OH &  0.896 & & {\mathrm{expt.}} & 18 \\
NH$_3$ & 1.476\pm .002 & & {\mathrm{expt.}} &  19 \\
H$_2$O & 1.847 & & {\mathrm{expt.}} &  20 \\
SiC$_2$ & 2.393\pm .006 & & {\mathrm{expt.}} &  21  \cr 
HCl & 1.109 & & {\mathrm{expt.}} &  22  \cr 
OH & 1.655 & & {\mathrm{expt.}} &  23  \cr 
 H$_3$O$^{+}$ & 1.44 &&  {\it ab\ initio}  &  24  \cr 





    \noalign{\smallskip}
            \hline
 \noalign{\smallskip}  
         \end{array}
          \smallskip
$$
 \noindent 
 $^{\mathrm a}$ Same data are also adopted for isotopes and deuterated
 species, unless stated in the datafile. 
 
  \noindent 
 $^{\mathrm b}$ All values are in units of
debye (D), where 1 D = $10^{-18}$ esu cm. When the original source has
presented $\mu$ in atomic units, a conversion factor of 1 au = $ea_0$ =
2.54175 D has been applied.

 \noindent
 Refs. -- (1) \citet{Goorvitch94}. (2) \citet{Lovas92}. (3) \citet{Patel79}. (4) \citet{Winnewisser68}.
 (5) \citet{Raymonda70}. (6) \citet{Hoeft69}. (7) \citet{Botschwina93}. (8) \citet{Botschwina89}.
 (9) \citet{DeFrees82}. (10) \citet{Muenter68}. (11) \citet{DeLeon85}. (12) \citet{Ebenstein84}.
 (13) \citet{Blackman76}. (14) \citet{Gadhi95}. (15) \citet{Fabricant77}. (16) \citet{Green74b}.
 (17) \citet{Botschwina85}. (18) \citet{Sastry81}. (19) \citet{Cohen74}. (20) From JPL based on \citet{Camy-Peyret85}. 
 (21) \citet{Suenram89}. (22) \citet{DeLeeuw71}.  (23) \citet{Peterson84}. 
 (24) \citet{Botschwina84}.
\end{table}


\section{Collisional rate coefficients}
\subsection{General considerations}
\label{collrates_general}

The rate of collision is equal to
\begin{equation}
C_{ul} = n_{\mathrm{col}} \gamma_{ul},
\end{equation}
where $n_{\mathrm{col}}$ is the number density of the collision
partner and $\gamma_{ul}$ is the downward collisional rate coefficient (in
cm$^3$\,s$^{-1}$).  The rate coefficient is the Maxwellian average of
the collision cross section, $\sigma$,
\begin{equation}
\label{ratecoeff}
\gamma_{ul} = \left(\frac{8kT}{\pi\mu}\right)^{-1/2}\left(\frac{1}{kT}\right)^2\int \sigma E e^{-E/kT} dE,
\end{equation}
where $k$ is the Boltzmann constant, $\mu$ is the reduced mass of
the system, and $E$ is the center-of-mass collision energy.  The upward rates are obtained through detailed balance
\begin{equation}
\label{balance}
\gamma_{lu} = \gamma_{ul} \frac{g_u}{g_l} e^{-h\nu/kT_{\mathrm{kin}}},
\end{equation}
where $g$ is a statistical weight.

The collisional rate coefficients $\gamma_{ul}$ usually pose
the largest source of uncertainty of the molecular data input to the
radiative transfer analysis (however, see discussion on dipole moments
in Sect.~\ref{dipole}). The dominant collision partner is often 
H$_2$ except in photon dominated regions (PDRs) where collisions with
electrons and H can become important. The collisional rate
coefficients presented here are mainly with H$_2$ and only in a few
cases (in particular the atoms) are collisions with H and electrons also treated.  Where
available, the data files include collisions with ortho- and
para-H$_2$, e.g., in the case of CO.

If only data for collisions with He are available, a first order correction can
be made by assuming H$_2$ to have the same cross sections. This
approximation is strictly only valid for very cold sources, where most
H$_2$ is in the ground $J=0$ state without angular momentum.
Then from Eq.~(\ref{ratecoeff}) the rate coefficient for
collisions between a molecular species X and H$_2$
\begin{equation}
\label{scale}
\gamma_{\mathrm{X-H_2}} = \gamma_{\mathrm{X-He}} \left( 
\frac{\mu_{\mathrm{X-He}}}{\mu_{\mathrm{X-H_2}}} \right)^{1/2}.
\end{equation}
If the mass of the molecule is much larger than that of He and H$_2$,
the scaling factor is 1.4.  

Some molecules of significant
interest lack calculated collisional rate coefficients. In these cases the rates
for a similar molecule have been adopted and only scaled for the
difference in reduced mass following Eq.~(\ref{scale}). This procedure
works best for O$\to$S substitutions (for example, scaling HCO$^+$
rates for the case of HCS$^+$) since such molecules have a similar molecular structure.

For most species, only rate coefficients with He or H$_2$ $J$=0 are
available. Values with H$_2$ $J$=1 can be larger by factors of 2--5
due to supplementary terms in the interaction potential (e.g. Green
1977, H$_2$O example). This additional uncertainty is often not
considered in astrophysical analyses.  In the case of CO and H$_2$O, separate
rate coefficients are available for collisions with ortho- and
para-H$_2$. The online version of RADEX weighs these coefficients by
the thermal value of the H$_2$ o/p-ratio at the kinetic temperature.
The o/p-ratio is approximated as the $J$=1 to $J$=0 population ratio with
a maximum of 3.0, which is an overestimate by at most 20\% (at
$T$=155~K). 
In the datafiles available for download the collisional rate
coefficients for collisions with ortho-H$_2$ and para-H$_2$ are kept
separate.

To obtain the collision rate, RADEX simply multiplies the collisional
rate coefficients with the H$_2$ density. To include the effect of
collisions with He, the user must multiply the density  by 1.14 (to first order) for a He
abundance with respect to H$_2$ of 20\%.

The adopted collisional rate coefficients are presented in Tables~\ref{tab_atoms}  and  \ref{tab_coll} for atomic and molecular species, respectively.   For isotopomers the same set of
collisional rate coefficients as for the main isotope was adopted, unless otherwise stated.
Tables~\ref{tab_atoms}  and  \ref{tab_coll} show the temperature range and 
maximum energy ($E_{\mathrm{max}}$) for which calculations are available. 
Also, the collision partner is indicated.
Only the downward values are given in the data
files; the upward rate coefficients are obtained through detailed
balance using Eq.~(\ref{balance}). 


%
   \begin{table}
      \caption[]{Summary of atomic collisional data from the literature.}
 \label{tab_atoms} 
 $$
         \begin{array}{p{0.2\linewidth}ccccccccc}    
    \hline
    \noalign{\smallskip}
    \multicolumn{1}{c}{\mathrm{Atom}} & 
            \multicolumn{1}{c}{T} && 
    \multicolumn{1}{c}{E_{\mathrm{max}}} &&
    \multicolumn{1}{c}{\mathrm{Collision}} &&
    \multicolumn{1}{c}{\mathrm{Ref.}}      \\
             &
             \multicolumn{1}{c}{\mathrm{[K]}} && 
    \multicolumn{1}{c}{[\mathrm{cm}^{-1}]} && \mathrm{partner}
             &&
             \\

    \noalign{\smallskip}
            \hline
            \noalign{\smallskip}
C               & \phantom{00}10-\phantom{00}200        &&    \phantom{000}43     && \mathrm{H} && 1 \\ 
                  & \phantom{00}10-20000                            &&    \phantom{000}43     && \mathrm{e}^{-} && 2 \\ 
                  & \phantom{0}100-\phantom{0}2000        &&    \phantom{000}43     && \mathrm{H}^{+} && 3 \\ 
                  & \phantom{00}10-\phantom{00}150        &&    \phantom{000}43     && \mathrm{He} && 4 \\ 
                  & \phantom{00}10-\phantom{0}1200        &&    \phantom{000}43     && \mathrm{H}_2 && 5 \\
C$^{+} $    &  \phantom{000}5-\phantom{0}3162      &&   \phantom{000}63     && \mathrm{H} && 1 \\  
                   &  \phantom{00}10-20000                          &&   43054     && \mathrm{e}^{-} && 6\\ 
                   & \phantom{00}10-\phantom{00}250        &&    \phantom{000} 63     && \mathrm{H}_2 && 7 \\
O               & \phantom{00}50-\phantom{0}1000        &&     \phantom{00}227     && \mathrm{H}      && 1 \\
                  & \phantom{00}50-\phantom{0}3000        &&     \phantom{00}227     && \mathrm{e}^{-} && 8 \\
                  & \phantom{0}100-\phantom{00}100        &&     \phantom{00}227     && \mathrm{H}^{+} && 9 \\
                  & \phantom{00}20-\phantom{0}1500        &&     \phantom{00}227     && \mathrm{H}_2 && 10 \\

    \noalign{\smallskip}
            \hline
 \noalign{\smallskip}  
         \end{array}
 $$
          \smallskip
 \noindent
 Refs. -- (1) \citet{Launay77}. (2) \citet{Johnson87}. (3) \citet{Roueff90b}. (4) \citet{Staemmler91}.
 (5) \citet{Schroeder91}.  (6) \citet{Wilson02}. (7) \citet{Flower77}.  (8) \citet{Bell98}. (9) \citet{Chambaud80}. (10) \citet{Jaquet92}.
   \end{table}
   \begin{table}
         \caption[]{Summary of molecular collisional data from the literature and new extrapolated rate coefficients.}
 \label{tab_coll} 
 $$  
  \begin{array}{p{0.2\linewidth}ccccccccc}

    \hline
    \noalign{\smallskip}
    \multicolumn{1}{c}{{\mathrm{Molecule}}} & 
            \multicolumn{1}{c}{T} && 
    \multicolumn{1}{c}{E_{\mathrm{max}}} &&
    \multicolumn{1}{c}{{\mathrm{Collision}}} &&
    \multicolumn{1}{c}{{\mathrm{Ref.}}}      \\
             &
             \multicolumn{1}{c}{\mathrm{[K]}} && 
    \multicolumn{1}{c}{[\mathrm{cm}^{-1}]} && \mathrm{partner}
             && 
              \\
                            
                  \noalign{\smallskip}
            \hline
            \noalign{\smallskip}
CO              & \phantom{00}5-\phantom{0}400        &&    1557     && \mathrm{H}_2 && 1 \\
                    &           100-2000                 &&   \phantom{0}729     && \mathrm{H}_2 && 2 \\  
                   &  \phantom{00}5-2000                 &&   3137     && \mathrm{H}_2 &&  3\\ 
                    & \phantom{00}5-2000                 &&   3137     && \mathrm{H}_2 && {\mathrm{this\ work}}^* \\ 
SO               &  \phantom{0}50-\phantom{0}350      && \phantom{0}405 && \mathrm{H}_2 && 4^* \\ 
SO$_2$       & \phantom{0}25-\phantom{0}125          && \phantom{00}62 && \mathrm{He}   && 5 \\ 
                   & \phantom{0}10-\phantom{0}375          && \phantom{0}250 && \mathrm{H}_2   && {\mathrm{this\ work}}^* \\ 
CS              & \phantom{0}20-\phantom{0}300       &&  \phantom{0}310 && \mathrm{H}_2  && 6 \\ 
                  & \phantom{0}20-2000       &&  1949 && \mathrm{H}_2  &&  {\mathrm{this\ work}}^*  \\ 
SiO              & \phantom{0}20-\phantom{0}300      && \phantom{0}275 && \mathrm{H}_2   && 6  \\ 
                  & \phantom{0}20-2000      && 1185 && \mathrm{H}_2   && {\mathrm{this\ work}}^*  \\ 
SiS             & \phantom{0}20-2000      && \phantom{0}496 && \mathrm{H}_2   && {\mathrm{this\ work}}^*  \\ 
HCO$^+$        & \phantom{0}10-\phantom{0}400        && \phantom{0}565 && \mathrm{H}_2   && 7  \\ 
                  & \phantom{0}10-2000      && 1381 && \mathrm{H}_2   && {\mathrm{this\ work}}^*  \\ 
OCS            & \phantom{0}10-\phantom{0}150          &&  \phantom{0}165 && \mathrm{He}     && 8  \\ 
                   & \phantom{0}10-\phantom{0}100          && \phantom{00}27 && \mathrm{H}_2   && 9  \\ 
                  & \phantom{0}10-2000      && \phantom{0}517 && \mathrm{H}_2   && {\mathrm{this\ work}}^*  \\ 
HC$_3$N        & \phantom{0}10-\phantom{00}80          &&   \phantom{00}64 && \mathrm{He}    && 9 \\
                  & \phantom{0}10-2000      &&  \phantom{0}387 && \mathrm{H}_2   && {\mathrm{this\ work}}^*  \\ 
HCN               & \phantom{00}5-\phantom{0}100          &&  \phantom{00}83 && \mathrm{He}    && 10  \\
  & 100-1200          &&  1284 && \mathrm{He}    && 11  \\ 
    & \phantom{0}10-\phantom{00}30          &&  \phantom{00}30 && \mathrm{He}    && 12  \\ 
                  & \phantom{00}5-1200      && 1284 && \mathrm{H}_2   && {\mathrm{this\ work}}^*  \\ 
HNC          & \phantom{00}5-1200      && \phantom{0}635 && \mathrm{H}_2   && {\mathrm{this\ work}}^*  \\ 
C$_3$H$_2$     &  \phantom{0}10-\phantom{00}30          &&   \phantom{00}82 && \mathrm{He}    && 13  \\ 
               &  \phantom{0}30-\phantom{0}120          &&   \phantom{00}82 && \mathrm{He}    && 14 \\
                                 & \phantom{0}10-\phantom{0}120      && \phantom{00}82 && \mathrm{H}_2   && 15^*  \\ 
H$_2$CO         & \phantom{0}10-\phantom{0}300 &&  \phantom{0}208 && \mathrm{He}   && 16^* \\
N$_2$H$^{+}$   &  \phantom{00}5-\phantom{00}40  &&  \phantom{00}47 && \mathrm{He}    && 17\\
                         &  \phantom{0}10-1000  &&  \phantom{0}652 && \mathrm{H}_2   && {\mathrm{this\ work}}^* \\
HCS$^{+}$&  \phantom{0}10-\phantom{00}60  &&  \phantom{00}64 && \mathrm{He}    && 18 \\
& \phantom{0}10-1000      && \phantom{0}360 && \mathrm{H}_2   && {\mathrm{this\ work}}^*  \\ 
CH$_3$OH      &  \phantom{00}5-\phantom{0}200      &&   \phantom{0}362  && \mathrm{H}_2  && 19^* \\
                           &  \phantom{00}5-\phantom{0}200      &&   \phantom{0}362  && \mathrm{He}  && 20,21  \\
NH$_3$      & \phantom{0}15-\phantom{0}300 && \phantom{0}416 && \mathrm{H}_2  && 22^*  \\
H$_2$O        & \phantom{0}20-2000 && 1395 && \mathrm{He}  && 23^* \\
                     & \phantom{00}5-\phantom{00}20 &&  && \mathrm{H}_2  && 24, 25 \\
                     & \phantom{0}20-\phantom{0}140 &&  && \mathrm{H}_2  && 26 \\
HDO             & \phantom{0}50-\phantom{0}500 &&  && \mathrm{He}  && 27^*  \\
SiC$_2$       & \phantom{0}25-\phantom{0}125 && \phantom{00}50 && \mathrm{H}_2  &&  15^*\\
OH                 & \phantom{0}15-\phantom{0}200 &&  \phantom{0}400  && \mathrm{H}_2    &&  28^* \\
HCl                & \phantom{0}10-\phantom{0}300 &&  \phantom{0}583 && \mathrm{He}    && 29^*  \\
H$_3$O$^+$  & 100-100 &&  \phantom{0}259 && \mathrm{H}_2    && 30^*  \\

            \hline
   \end{array}
          $$
$^*$ Datafile adopted in the online version of RADEX\\
 \noindent
 Refs. -- (1) \citet{Flower01a}. (2) \citet{Schinke85}. (3) \citet{Larsson02}. (4) \citet{Green94}. (5) \citet{Green95}. (6) \citet{Turner92}. (7) \citet{Flower99}.
          (8) \citet{Flower01b}. (9) \citet{Green78a}. (10) \citet{Green74}. (11) Green (unpublished data). (12) \citet{Monteiro86}. (13) \citet{Avery89}. (14) \citet{Green87}. (15) \citet{Chandra00}. (16) \citet{Green91}. (17) \citet{Green75b}. 
 (18) \citet{Monteiro84}. (19) \citet{Pottage04}. (20) \citet{Pottage01}. (21) \citet{Pottage02}.  (22) \citet{Danby88}. (23) \citet{Green93}. (24) \citet{Dubernet02}. (25) \citet{Grosjean03}. (26) \citet{Phillips96}. (27) \citet{Green89}. (28) \citet{Offer94}. (29) \citet{Neufeld94}. (30) \citet{Phillips92}. 
   \end{table}

\subsection{Accuracy of adopted rate coefficients}

Most of the collisional data summarized in Table 2 have been obtained
from theoretical calculations, with experimental cross-checks possible
for only a few cases. Most experiments reflect the average of many
collisional events, with comparisons typically done for relaxation
rates and collision-induced pressure line broadening.  State-to-state
measurements have been possible for only a few systems and they often
report relative rather than absolute cross sections.  Also,
experiments with H$_2$ are usually done for n-H$_2$ (i.e., 75\%
o-H$_2$ and 25\% p-H$_2$), rather than for H$_2$ $J$=0 or 1.
Nevertheless, such comparisons between theory and experiment, as well
as those between different theoretical methods, have given some
indication of the uncertainties in the collisional rate
coefficients. Excellent accounts of the methods involved and details
on individual systems are given by \citet{Green75}, \citet{Flower90} and
\citet{Roueff90}; recent developments are reviewed by
\citet{roueff04}. Here only a brief summary is given. 

The theoretical determination of collisional rate coefficients
consists of two steps: (i) determination of the interaction potential
$V$ between the colliding systems; and (ii) calculation of the
collision dynamics.  Significant progress in the second part has been
made in the last decades, aided by the increased computer speed. The
most accurate method is the Close-Coupling (CC) method, in which the
scattering wave function is expanded into a set of basis
functions. This method is exact if an infinite number of basis
functions or `channels' is taken into account. In practice a finite
number of channels is used, resulting in a set of coupled second-order
differential equations. The absolute accuracy of the results can
easily be checked by increasing the basis set, and is of order a few
\% for a given interaction potential.  This method works very well for
low collision energies and relatively light species, although care
should be taken at the lowest energies whether resonances are properly
sampled \citep[e.g.,][]{Dubernet02}. However, the method becomes
increasingly computationally demanding at high energies and for heavy
polyatomic molecules with small splittings between the rotational
energy levels resulting in many channels.

The most popular approximate dynamical methods are the Coupled States
(CSt) or `centrifugal decoupling' method and the Infinite Order Sudden
(IOS) approximation. In the CSt method, the centrifugal potential is
assumed to conserve the projection of the angular momentum on the axis
perpendicular to the plane of the collision partners.  This
approximation is often valid at higher energies if the collision is
dominated by the repulsive part of the potential. In the IOS
approximation, an additional assumption is that the molecule does not
rotate during collisions. This may be appropriate for heavy rotors at
energies much larger than the rotational energies. From comparisons
with the more exact CC results, it is found that absolute
uncertainties for the CSt method range from $\sim$10\% to a factor of
2, with lesser uncertainties in the relative values. The propensities
in the collisions are recovered correctly. In contrast, the IOS method
can have uncertainties up to an order of magnitude.  Computer programs
which include the CC, CSt and IOS options are publically available
(see \nocite{Hutson94}Hutson \& Green 1994\footnote{\tt
http://www.giss.nasa.gov/molscat}, \nocite{Flower00}Flower et al.\ 2000\footnote{\tt
http://ccp7.dur.ac.uk/molcol.html}, Manolopoulos 1986\nocite{Manolopoulos86} and Alexander \& Manolopoulos 1987\nocite{Alexander87}\footnote{{\tt
http://www.chem.umd.edu/physical/alexander/hibridon}}

The above quoted ranges of uncertainties assume that the interaction
potential is perfectly known. Often, this is not the case and the
potential surfaces form the largest source of error in the collisional
rates with uncertainties that are difficult to assess. The interaction
potential consists of a short-range repulsive part, an
intermediate-range interaction part where a weak molecular bond is
formed, and a long-range part dominated by electrostatic
interaction. The intermediate part is most difficult to determine and
requires high-level quantum chemical models. The most accurate method
is that of Configuration Interaction (CI), but it can become very
costly in computer time. Other methods include Hartree-Fock
Self-Consistent-Field (SCF) and perturbation methods, and more
recently Density Functional Theory (DFT), but each of these methods
has its drawbacks. An old approximate method, the Electron Gas model,
is now obsolete, but some dynamics calculations for astrophysical
systems still use these potentials 
(e.g., CS-H$_2$, \nocite{Turner92}Turner et al.\ 1992).

The following selected examples serve to illustrate the range of
absolute errors in the adopted collisional rate coefficients.  It
should be noted that relative values often have less uncertainty and
that these are most relevant for astrophysical applications: small
absolute errors can often be  compensated by small adjustments in the
abundance of the species.  

\begin{figure}
\centerline{\includegraphics[height=7cm, angle=-90]{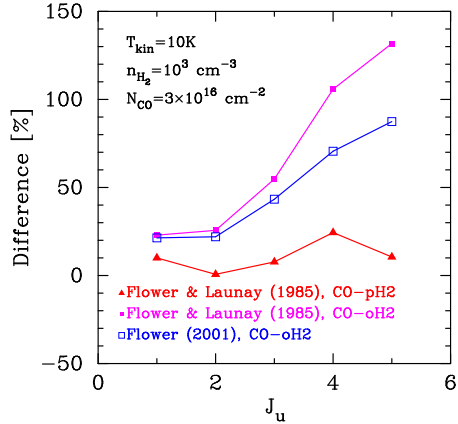}}
  \caption{Predicted CO line intensities, using different sets of calculated
 collisional rate coefficients, for an isothermal homogeneous sphere with
 a kinetic temperature 10\,K, a H$_2$ density of $10^3$\,cm$^{-3}$ and a CO 
 column density of  $3\times10^{16}$\,cm$^{-2}$.The line intensities are shown in relation to the values
    obtained using the CO-pH$_2$ rate coefficients from
    \citet{Flower01a}.   The upper rotational quantum number $J_{\mathrm{u}}$ is indicated on the $x$-axis.      
     The rotational transitions are out of thermal equilibrium  and, for transitions below $J=4\rightarrow3$, 
 optically thick. }
 \label{co_mod}      
\end{figure}

\subsubsection{CO--H$_2$} 
Early calculations by \citet{Green76},
\citet{Schinke85} and \citet{Flower85} illustrate the
sensitivity of the results to different potential energy
surfaces. Absolute differences in individual collisional rate
coefficients range from a few \% up to 40\%, with the relative values
usually having less scatter.  Comparison of computed cross sections
using a new CO--H$_2$ potential by \citet{Jankowski98}
with pressure broadening and scattering experiments by \citet{Mengel00b}
 suggests an overall average absolute accuracy of better than 10\%
at $T\geq 30$K, but somewhat less good at the lowest temperatures
where the deviations can increase to 30--50\%.  No information is
available on the accuracy of the larger $\Delta J$ transitions (e.g.,
$\Delta J>10$), which become important at high temperatures such as
found in dense shocks. The same potential surface has been used in the latest set of rate coefficients given by \citet{Flower01a} which are adopted here.

The following simple test problem illustrates the consequences of using different sets of collisional rate coefficients. Line intensities were calculated for
the lowest 5 rotational transitions of CO  for a molecular cloud of constant temperature 
and density using RADEX.  The model has a temperature of 10~K,  H$_2$ density of $1\times10^3$~cm$^{-3}$ and a total CO column density of $3\times10^{16}$~cm$^{-2}$  over a
line width (full-width at half-maximum) of 1 km s$^{-1}$. All lines are out of thermal equilibrium and the three lowest rotational transitions are optically thick.  As is shown in Fig.~\ref{co_mod}, differences of up to $\pm 150$\% are found, especially  for collisions with para-H$_2$
compared with ortho-H$_2$.


\subsubsection{H$_2$CO--H$_2$} 
The H$_2$CO--H$_2$ rate coefficients given in our
database are obtained from \citet{Green91}, who calculated values for
the H$_2$CO--He system using a very old potential energy surface by
\citet{Garrison75} based on SCF and limited CI calculations.
These rate
coefficients and the adopted surface have  recently been tested against
pressure broadening and time-resolved double-resonance studies for
three low-lying transitions \citep{Mengel00}. Satisfactory
agreement is found for the H$_2$CO--He system, with differences in
cross sections ranging from a few \% up to 20\%.  The deviations are
largest at the lowest temperatures, $<$10~K, as was also found for
CO--H$_2$. For H$_2$ as the collision partner, the cross sections are
found to be up to a factor of two higher, significantly more than the
value of 1.4 expected from the difference in masses, illustrating that
simple scaling from He collisions may introduce errors up to 50\%.

%

\subsubsection{OH--H$_2$} 
One of the computationally most challenging systems
is OH--H$_2$, since OH is an open shell molecule with a $^2\Pi$ ground
state so that two potential surfaces and hyperfine splitting are
involved. Results for collisions with both o-H$_2$ and p-H$_2$ are
presented by \citet{Offer92} using an old potential
surface based on SCF calculations \citep{Kochanski81}, and by
\citet{Offer94} using a new surface computed using CI \citep{Offer93}. 
The differences due to the potential energy surface
range from 10 \% to more than an order of magnitude for individual
rate coefficients. Comparison with state-to-state experimental cross
sections with both n-H$_2$ and p-H$_2$ at one specific energy gives
surprisingly good agreement, usually within 50\% but with occasional
excursions up to an order of magnitude \citep{Schreel96}.
Moreover, all the propensities for individual hyperfine transitions
are well reproduced.

\begin{figure*}
\centerline{\includegraphics[width=14cm]{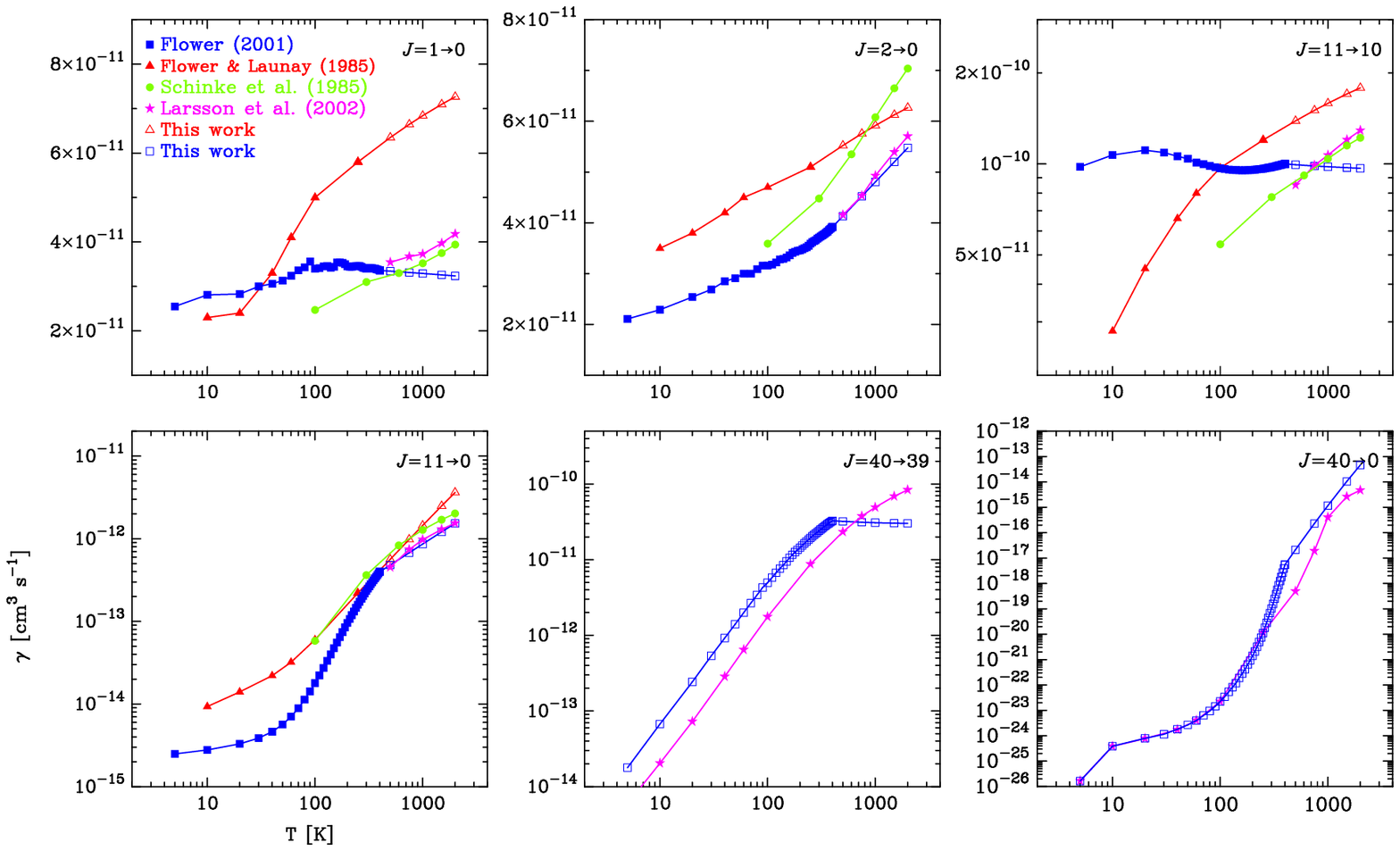}}  
  \caption{Calculated and extrapolated collisional de-excitation rate coefficients for
  CO in collisions with para-H$_2$. Open triangles indicate extrapolation in temperature to the
  rate coefficients of \citet{Flower85} (filled triangles). Open squares show the extrapolation to higher temperatures and energy levels of the recent rate coefficients calculated by \citet{Flower01a} (filled squares).  For comparison the rate coefficients presented by \citet{Schinke85} (filled circles) and the extrapolation performed by \citet{Larsson02} (filled stars) are shown.}
  \label{fig_co}
\end{figure*}

\subsection{Adopted collisional rate coefficients}
Below follows a summary of the collisional rate coefficients adopted
in the first release of the database. 
Molecules for which only one set of calculated collisional rate
coefficients is available and where no extrapolation was performed
are not described further here.  The principle method for extrapolating the downward 
collisional rate coefficients  ($\Delta
J$$=$$J_{u}$$\rightarrow$$J_{l}$, 
$J_{u}$$>$$J_{l}$)  in temperature in the case of a linear molecule is 
\citep{deJong75,Bieging98}
\begin{equation}
\label{collrates2b}
\gamma_{ul} = a(\Delta J)y\exp[-b(\Delta J)y^{1/4}]\times\exp[-c(\Delta J)y^{1/2}],
\end{equation}
where $y=\Delta E_{ul}/kT$ and the three parameters $a$, $b$, and $c$
are determined by least-squares fits to the initial set of rate
coefficients for each $\Delta J$. This reproduces most of the calculated rate coefficients
to within 50\% and typically within 20\%.
For more details on the extrapolation
scheme, including extrapolation in energy levels, see Sect.~\ref{extrapolation}  (only available in the online version of this journal).

When the specified kinetic temperature falls outside the region where collisional rate
coefficients are available, i.e.\ from $T_{\mathrm{low}}$ to $T_{\mathrm{high}}$, RADEX makes no further extrapolation and assumes  the downward rate coefficients at $T_{\mathrm{low}}$ and $T_{\mathrm{high}}$, respectively.

\subsubsection{CO}
\label{adopted_CO}
For CO the collisional rate coefficients calculated by \citet{Flower01a} have 
been adopted as a starting point.  These computations cover 
 temperatures in the range from 5~K up to 400~K and
include rotational levels up to $J=29$ and $J=20$ for collisions with
para-H$_2$ and ortho-H$_2$, respectively. Both sets of rate
coefficients were then extrapolated to include energy levels up to
$J=40$  [using Eq.~(\ref{collrates1})] and temperatures up to 2000~K  [using Eq.~(\ref{collrates2b})] , as described in
Sect.~\ref{linear_collrates}.  In the datafile available for download,
the collisional rate coefficients for collisions with ortho-H$_2$ and
para-H$_2$ are kept separate. However, in RADEX they are weighted
together as described in Sect.~\ref{collrates_general}.

Fig.~\ref{fig_co} shows the extrapolation of CO collisional de-excitation
rate coefficients for collisions with para-H$_2$. 
It is clear that extrapolated rate
coefficients are uncertain and depend on both the original data set
from which the extrapolation is made and the method adopted. However,
the extrapolated values  typically agree within 50\% in the case of CO.
The largest discrepancies, up to an order of magnitude, naturally
arise in the region where extrapolation in both temperature and energy
levels are performed.
Thus, in the parts of parameter space where extrapolated rates are being used to infer  
physical conditions, care should be taken as to any astrophysical conclusions drawn 
from the modeling.

\subsubsection{CS}
For CS the rate coefficients calculated by \citet{Turner92} have been
adopted as a starting point.  These values have been computed 
 for temperatures in the range $20- 300$~K and include
rotational levels up to $J=20$ for collisions with H$_2$. This set 
 was then extrapolated to include energy levels up
to $J=40$  [using Eq.~(\ref{collrates1})] and temperatures up to 2000~K
 [using Eq.~(\ref{collrates2b})], as described in
Sect.~\ref{linear_collrates}. No extrapolation to  temperatures
lower  than 20~K was attempted.

\subsubsection{SiO}
For SiO the rate coefficients calculated by \citet{Turner92} have been
adopted, computed for temperatures in the range $20-300$~K and including
rotational levels up to $J=20$ for collisions with H$_2$. This set 
was then extrapolated to include energy levels up
to $J=50$  [using Eq.~(\ref{collrates1})] and temperatures up to 2000~K  [using Eq.~(\ref{collrates2b})], as described in
Sect.~\ref{linear_collrates}. No extrapolation to temperatures lower
than 20~K was attempted.

\subsubsection{SiS}
No calculated rate coefficients are available for SiS. Instead, the
same set of collisional rate coefficients as for SiO has been adopted.

\subsubsection{HCO$^+$}
The rate coefficients for HCO$^+$ in collisions with H$_2$ have been calculated by
 \citet{Flower99} for temperatures in the range $10-400$~K and
  rotational levels up to $J=20$. This set of rate
coefficients was then extrapolated to include energy levels up to
$J=30$  [using Eq.~(\ref{collrates1})] and temperatures up to 2000~K  [using Eq.~(\ref{collrates2b})], as described in
Sect.~\ref{linear_collrates}.

\subsubsection{HC$_3$N}
The rate coefficients for HC$_3$N in collisions with He
have been calculated by \citet{Green78a} for temperatures in the range $10-80$~K and 
rotational levels up to $J=20$. This set of rate coefficients was then
extrapolated to include energy levels up to $J=50$  [using Eq.~(\ref{collrates1})] and
temperatures up to 2000~K  [using Eq.~(\ref{collrates2b})], as described in
Sect.~\ref{linear_collrates}. The rate coefficients were then scaled
by 1.39 to represent collisions with H$_2$ instead of He.

\subsubsection{HCN}
The rate coefficients for HCN in collisions with He
have been calculated by \citet{Green74} for temperatures in the range $5-100$~K and 
rotational levels up to $J=7$. This work has subsequently been extended by S. Green (unpublished data) to include rotational levels up to $J=29$ and temperatures from $100-1200$~K. Extrapolation of the rate coefficients  to include energy levels up to
$J=29$ for temperatures below 100~K  [using Eq.~(\ref{collrates1})], as described in
Sect.~\ref{linear_collrates}, has been made. The rate coefficients were subsequently
scaled by 1.37 to represent collisions with H$_2$ instead of He. 
 The collisional rate coefficients between various
hyperfine levels have been calculated by \citet{Monteiro86} for the lowest ($J\leq4$) rotational levels
and temperatures from $10-30$~K in collisions with He. A datafile
based on these collisional rate coefficients  is
also made available separately.

\subsubsection{HNC}
No calculated rate coefficients are available for HNC. Instead, the
same set of collisional rate coefficients as for HCN has been adopted.

\subsubsection{N$_2$H$^+$}
The rate coefficients for N$_2$H$^+$ in collisions with He
atoms have been calculated by \citet{Green75b} for temperatures in the range $5-40$~K and 
rotational levels up to $J=6$. Given the limited range in temperature
and energy levels, we have instead adopted the same rate
coefficients as for HCO$^+$. This is motivated by the discussion in
\citet{Monteiro84}  where the rate coefficients for these two species in collisions with He are found to be very similar, typically within 10\%. 

\subsubsection{HCS$^+$}
The rate coefficients for HCS$^+$ in collisions with He
atoms have been calculated by \citet{Monteiro84} for temperatures in the 
range $10-60$~K and rotational levels up to $J=10$. This set of rate coefficients was then
extrapolated to include energy levels up to $J=23$  [using Eq.~(\ref{collrates1})] and
temperatures up to 1000~K  [using Eq.~(\ref{collrates2b})], as described in
Sect.~\ref{linear_collrates}. The rate coefficients were subsequently scaled
by 1.38 to represent collisions with H$_2$ instead of He.

%

\subsubsection{H$_2$O}
In RADEX the rate coefficients for H$_2$O in collisions with He calculated by \citet{Green93} 
are used as default. The rates were computed for temperatures in the range from 20 to 2000~K including energy levels up to about 1400~cm$^{-1}$. These rate coefficients were subsequently scaled
by 1.35 to represent collisions with H$_2$ instead of He.
In addition, a datafile containing the recent rate coefficients for  H$_2$O in collisions with  p-H$_2$ \citep{Grosjean03} and 
o-H$_2$ \citep{Dubernet02} calculated for low temperatures ($5-20$~K) has been constructed.
In the datafiles available for download,
the rate coefficients for collisions with ortho-H$_2$ and
para-H$_2$ are kept separate. However, in RADEX they are weighted
together as described in Sect.~5.1.

\subsubsection{SO$_2$}
 For non-linear molecules there are no simple scaling relations such as Eq.~(\ref{collrates2b}). 
In Sect.~6.2 (only available in the online version of this paper) the procedure adopted to extrapolate  rate coefficients for SO$_2$ is presented. As starting point the calculated rate coefficients for SO$_2$ in collisions with He calculated by \citet{Green95} were used. These rates were computed for temperatures in the range from 25 to 125~K including energy levels up to about 62~cm$^{-1}$.
Extrapolation  was made to include energy levels up to 250~cm$^{-1}$ and
temperatures in the range from 10 to 375~K. The rate coefficients were subsequently scaled
by 1.4 to represent collisions with H$_2$ instead of He.

\section{Summary}

A compilation of atomic and  molecular data  in
a homogeneous format relevant for radiative transfer modelling is presented. The
data files are made available through the WWW and include energy
levels, statistical weights, Einstein A-coefficients and collisional
rate coefficients. Extrapolation of collisional rate coefficients are
generally needed and different schemes for this are reviewed.

In addition to the atomic and molecular database,  an online version of a computer code for performing
statistical equilibrium calculations is made available for use through
the WWW. The program, named RADEX, is an alternative to the widely
used rotation diagram method and has the advantage of supplying the
user with physical parameters such as density and temperature.

Databases such as these depend heavily on the efforts by the chemical physics community to provide the relevant atomic and molecular data. We strongly encourage further efforts in this direction, so that the current extrapolations of collisional rate coefficients can be replaced by actual calculations in future releases.

\begin{acknowledgements}
  The authors are grateful to D.J.\ Jansen and F.P.\ Helmich for
  contributions to the data files and programs.  B. Larsson is thanked
  for providing his collisional rate coefficients for CO. 
   The referee A.~Markwick is thanked for a constructive report that helped improve both the paper as well as the online database.
  This
  research was supported by the Netherlands Organization for
  Scientific Research (NWO) grant 614.041.004 and a NWO Spinoza grant.
  FLS and JHB further acknowledge financial support from the Swedish Research
  Council.
\end{acknowledgements}

\bibliographystyle{aa}


\Online

\section{Extrapolation of collisional rate coefficients}
\label{extrapolation}
\subsection{Linear molecules}
\label{linear_collrates}
An often adopted starting point when fitting and extrapolating
 collisional rate coefficients is to take advantage of the IOS approximation 
 in which the entire matrix of
 state-to-state rate coefficients can be calculated from the basic
 $\gamma_{L0}$ rate coefficients  \citep[e.g.][]{Goldflam77}
\begin{equation}
\label{collrates}
\gamma_{JJ^{\prime}} = (2J^{\prime}+1)\sum_{L=|J-J^{\prime}|}^{J+J^{\prime}} (2L+1)
\begin{pmatrix}
J & J^{\prime} & L \\
0 & 0 & 0  
\end{pmatrix}^{2}
\gamma_{L0},
\end{equation}
where 
\begin{equation}
\begin{pmatrix}
J & J^{\prime} & L  \\
0 & 0 & 0  
\end{pmatrix}
\end{equation}
is the Wigner 3-j symbol. This expression is valid only in the limit
where the kinetic energy of the colliding molecules is large compared
to the energy splitting of the rotational levels. Since the
energy splitting increases with $J$ this expression becomes less
accurate for higher rotational levels. \citet{DePristo79} show
that  by multiplying Eq.~(\ref{collrates}) (within the summation) with  
\begin{equation}
A(L,J)=\frac{6+\Omega(L)^2}{6+\Omega(J)^2},
\end{equation}
where 
\begin{equation}
\Omega(J^{\prime})=0.13\,J^{\prime} B_0\,l \left( \frac{\mu}{T}\right) ^{1/2},
\end{equation}
one can approximately correct for this deficiency. Here $B_0$ is
the rotational constant in cm$^{-1}$, $l$ is the scattering length in
{\AA} (typically $l \approx 3$ \AA), $\mu$ is the reduced mass of the
system in amu and $T$ is the kinetic temperature in K. 
Extrapolation of the rate coefficients down to the lowest $J$$=$0
level can be made both in temperature as well as in $J$ allowing the
general state-to-state coefficients to be extended \citep[e.g.,][]{Albrecht83,Larsson02}. 

\begin{figure}
\centerline{\includegraphics[height=8cm,angle=-90]{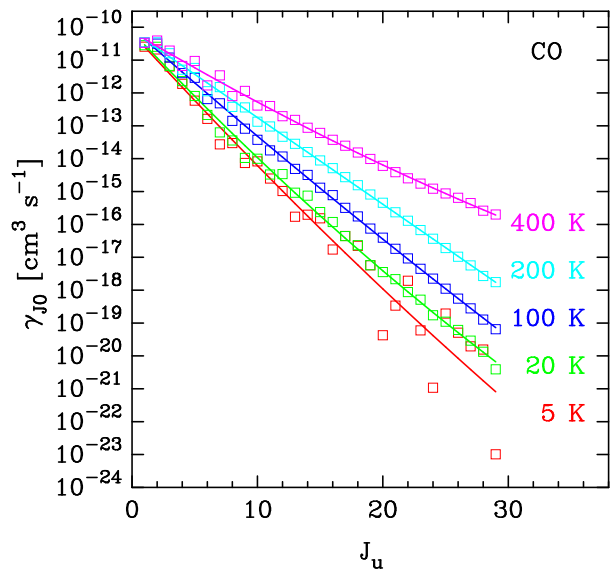}}
  \caption{The solid lines are fits to the CO--p-H$_2$ collisional rate coefficients from
 \citet{Flower01a} (squares ) for transitions down to the ground state from upper
 energy levels $J_{\mathrm u}$ using a second order polynomial. } 
 \label{co_iso}
\end{figure}

Alternatively, and in line with the IOS approximation, the downward
collisional rate coefficients ($\Delta
J$$=$$J_{u}$$\rightarrow$$J_{l}$, 
$J_{u}$$>$$J_{l}$) can be extrapolated in temperature using Eq.~(\ref{collrates2b}).
%
%
Given its simplicity we have
adopted this procedure for extrapolation of the rate coefficients in
temperature. 

Extrapolation to include higher rotational
levels was carried out by fitting the collisional rate coefficients connecting to the ground rotational state, at a
particular temperature, to a second order polynomial 
\begin{equation}
\label{collrates1}
\gamma_{J0} = \exp(a+bJ+cJ^2),
\end{equation}
where $a$, $b$ and $c$ are parameters determined from the fit. Figure~\ref{co_iso} illustrates the fit to collisional rate coefficients down to the ground rotational state for CO--H$_2$ using Eq.~(\ref{collrates1}).  
Similar extrapolations can be made in temperature. However, here we have adopted the approach by
\citet{deJong75} and \citet{Bieging98} and used Eq.~(\ref{collrates2b})
for the extrapolation in temperature. This extends the fit over a
larger range of energies. 
The IOS approximation [Eq.~(\ref{collrates})] was then used to calculate the entire matrix of
 state-to-state rate coefficients. The CO molecule is used in Sect.~\ref{adopted_CO} to illustrate the above mentioned schemes.

\subsection{Non-linear molecules}
\label{sec:app_so2}
For non-linear species there are no simple scaling relations and one has to
resort to custom-made fitting formulae for each case. The only case considered here is that
of SO$_2$ used by \citet{fvdt03}. In the prolongation of this project extrapolated collisional rate coefficients will be presented for additional non-linear molecules.
As the starting point for inelastic collisional data for \soo, the results of
\citet{Green95} were used. However, those data only cover the lowest 50
states, up to 62~\rcm\ ($J\approx 12$), while states up to $J=25$ are
commonly observed. 

\begin{figure}[t]
\resizebox{\hsize}{!}{\includegraphics{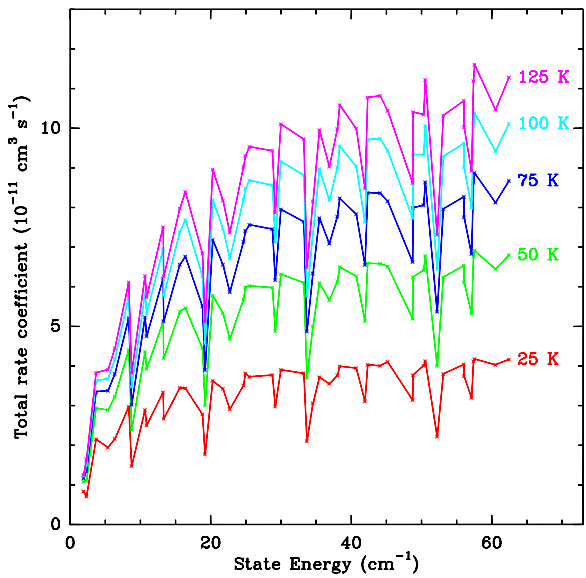}}  
  \caption{Collisional de-excitation rate coefficients for the lowest 50 
    states of \soo summed over all lower levels, 
  calculated from the data by Green (1995) for various temperatures.}

  \label{fig:so2_sums}
\end{figure}

Figure~\ref{fig:so2_sums} plots Green's downward rate coefficients,
summed over all final states, as functions of initial state. These
sums approach asymptotic values for $E_u \gtsim 40$\,cm$^{-1}$.
Deviations from this behaviour due to detailed quantum mechanical
selection rules are seen not to exceed 20\%.
The figure also shows that the rate coefficients increase
approximately as $T^{1/2}$, again to $\approx$20\% accuracy. This
behaviour indicates that the total rate coefficients only depend on
temperature through the collision velocity, while the de-excitation
cross sections are constant.

\begin{figure}[t]
\resizebox{\hsize}{!}{\includegraphics{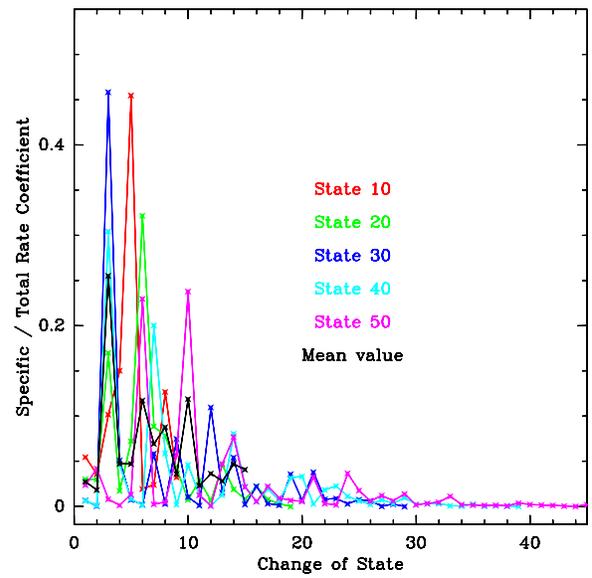}}    
  \caption{State-to-state de-excitation rate coefficients for SO$_2$
    as fractions of the total downward rate coefficient
    (Fig.~\ref{fig:so2_sums}), as a function of the number of levels
    by which the transition is changed.  The light (coloured) curves
    are values from Green (1995) at T=25~K for the 10$^{\rm th}$,
    20$^{\rm th}$, 30$^{\rm th}$, 40$^{\rm th}$ and 50$^{\rm th}$
    state above ground. The thick black curve is the normalized mean
    of the light (coloured) curves, adopted here to extrapolate
    Green's rate coefficients to higher-lying levels. The states are
    labelled in order of increasing energy.}
  \label{fig:so2_specs}
\end{figure}

Fig.~\ref{fig:so2_specs} shows that most collisions lead to
de-excitation into states that are not far down in energy. The
thick black curve is our fit to this behaviour: it is the normalized mean of
the various thin light curves which represent Green's data.  Transitions by
more than 15 states are considered negligible.

Based on these trends, the rate coefficients for de-excitation of \soo\ in inelastic
collisions with He are extrapolated as follows. For the 50 lowest
states, Green's values at $25<T<125$~K are used and multiplied by
$(T/125\,{\rm K})^{1/2}$ at temperatures up to 375\,K and down to
10\,K. For states between 62 and 250~\rcm\ above ground, a total
de-excitation rate coefficient of $1.0 \times 10^{-11}
T^{1/2}$\,\ccms\ is assumed, shown by Fig.~\ref{fig:so2_sums} to be a good
zeroth-order description for other levels. The state-to-state rate
coefficients are derived by multiplying these totals by the mean
propensities from \citet{Green95}, given by the black curve in
Fig.~\ref{fig:so2_specs}.  All results are multiplied by 1.4 to account
for the mass difference between H$_2$ and He. While this procedure is
admittedly crude and does not take the detailed quantum mechanics of
the interaction into account, it catches the spirit of more detailed
calculations.



\end{document}